\documentclass[aps,preprint,prb,showpacs,showkeys]{revtex4-1}%
\usepackage{amssymb}
\usepackage{amsfonts}
\usepackage{amsmath}
\usepackage{graphicx}%
\providecommand{\U}[1]{\protect\rule{.1in}{.1in}}

\begin{document}
\preprint{REV\TeX4-1 }
\title{On relationship between canonical momentum and geometric momentum}
\author{S. F. Xiao}
\affiliation{School for Theoretical Physics, School of Physics and Electronics, Hunan
University, Changsha 410082, China}
\affiliation{School of Physics Science and Technology, Lingnan Normal University, Zhanjiang
524048, China}
\author{Q. H. Liu}
\email{quanhuiliu@gmail.com}
\affiliation{School for Theoretical Physics, School of Physics and Electronics, Hunan
University, Changsha 410082, China}
\date{\today}

\begin{abstract}
Decompositing of $N+1$-dimensional gradient operator in terms of Gaussian
normal coordinates $(\xi^{0},\xi^{\mu})$, ($\mu=1,2,3,...,N$) and making the
canonical momentum $P_{0}$ along the normal direction $\mathbf{n}$ to be
hermitian, we obtain $\mathbf{n}P_{0}=-i\hbar\left(  \mathbf{n}\partial
_{0}-\mathbf{M}_{0}\right)  $ with $\mathbf{M}_{0}$ denoting the mean
curvature vector on the surface $\xi^{0}=const.$ The remaining part of the
momentum operator lies on the surface, which is identical to the geometric one.


\end{abstract}

\maketitle

\section{Introduction}

Exploration of the proper form and the physical meaning of momentum in
curvilinear coordinates attracts constant attention since the birth of quantum
mechanics. \cite{bp,RH,1968,LNF,DC,1989,2002,dong,GK,homma,ikegami,OK,liu2015}
In this paper, we show that canonical momenta $P_{\xi}$ associated with its
conjugate canonical positions, or coordinates, $\xi$, are closely related to
mean curvatures of the surface $\xi=const$. So, the geometric momenta
\cite{homma,ikegami,OK,liu2015,07,liu11,liu13-2,133,liu13-1,135,134,136,137,gem,WZ,iran,eprint}
that are under extensive studies and applications are closely related to
natural decompositions of the momentum operator in gaussian normal,
curvilinear in general, coordinates.

In next section II, we study a simple but illuminating example: how the
momentum operators in $3D$ spherical polar coordinates ($r,\theta,\varphi$)
are related to three mean curvature vectors. In section III, we present a
theorem for the general case. In final section IV, a brief conclusion is given.

\section{An example: mean curvatures and spherical polar coordinates}

The gradient operator in the $3D$ cartesian coordinate system $\nabla
_{cart}\equiv\mathbf{e}_{x}\partial_{x}+\mathbf{e}_{y}\partial_{y}%
+\mathbf{e}_{z}\partial_{z}$\ can be expressed in the $3D$ spherical polar
coordinates ($r,\theta,\varphi$),
\begin{equation}
\nabla_{sp}=\mathbf{e}_{r}\frac{\partial}{\partial r}+\mathbf{e}_{\theta}%
\frac{1}{r}\frac{\partial}{\partial\theta}+\mathbf{e}_{\varphi}\frac{1}%
{r\sin\theta}\frac{\partial}{\partial\varphi}. \label{grad}%
\end{equation}
The momentum operator can thus be written in the following way,
\begin{equation}
\mathbf{P}\mathbf{\equiv}-i\hbar\nabla_{cart}=-i\hbar\nabla_{sp}=\left\{
\mathbf{e}_{r},P_{r}\right\}  +\frac{1}{r}\left\{  \mathbf{e}_{\theta
},P_{\theta}\right\}  +\frac{1}{r\sin\theta}\left\{  \mathbf{e}_{\varphi
},P_{\varphi}\right\}  , \label{2}%
\end{equation}
where $\{A,B\}\equiv(AB+BA)/2$ and,
\begin{subequations}
\begin{align}
\left\{  \mathbf{e}_{r},P_{r}\right\}   &  =\mathbf{e}_{r}P_{r},P_{r}%
=-i\hbar(\frac{\partial}{\partial r}+\frac{1}{r}),\\
\{\mathbf{e}_{\theta},P_{\theta}\}  &  =\mathbf{e}_{\theta}P_{\theta}%
+i\hbar\frac{\mathbf{e}_{r}}{2},P_{\theta}=-i\hbar(\frac{\partial}%
{\partial\theta}+\frac{1}{2}\cot\theta),\\
\{\mathbf{e}_{\varphi},P_{\varphi}\}  &  =\mathbf{e}_{\varphi}P_{\varphi
}+i\hbar\frac{1}{2}\left(  \mathbf{e}_{r}\sin\theta+\mathbf{e}_{\theta}%
\cos\theta\right)  ,P_{\varphi}=-i\hbar\frac{\partial}{\partial\varphi}.
\end{align}
On one hand, these spherical polar coordinates have three mutually orthogonal
surfaces given by $r=const.$, $\theta=const.$ and $\varphi=const.$
respectively. They are, respectively, a\ spherical surface of radius $r$, a
cone of polar angle $\theta$, and a flat plane alone azimuthal angle $\varphi
$. These (curved) surfaces have three mean curvature vectors, respectively,
\end{subequations}
\begin{equation}
\mathbf{M}_{r}\mathbf{=}-\frac{\mathbf{e}_{r}}{r},\text{ }\mathbf{M}_{\theta
}\mathbf{=}-\frac{\mathbf{e}_{\theta}}{2r\tan\theta},\text{ }\mathbf{M}%
_{\varphi}\mathbf{=}0,\text{ }(r\neq0).
\end{equation}
On the other hand, if looking closely into the canonical momenta multiplied by
their vector coefficients, $\mathbf{e}_{r}$, $\mathbf{e}_{\theta}/r$ and
$\mathbf{e}_{\varphi}/\left(  r\sin\theta\right)  $, respectively, we find,
\begin{subequations}
\begin{align}
\mathbf{e}_{r}P_{r}  &  =-i\hbar\mathbf{e}_{r}(\frac{\partial}{\partial
r}+\frac{1}{r})=-i\hbar(\mathbf{e}_{r}\frac{\partial}{\partial r}%
-\mathbf{M}_{r}),\\
\frac{\mathbf{e}_{\theta}}{r}P_{\theta}  &  =-i\hbar\frac{\mathbf{e}_{\theta}%
}{r}(\frac{\partial}{\partial\theta}+\frac{1}{2}\cot\theta)=-i\hbar
(\frac{\mathbf{e}_{\theta}}{r}\frac{\partial}{\partial\theta}-\mathbf{M}%
_{\theta}),\\
\frac{\mathbf{e}_{\varphi}}{r\sin\theta}P_{\varphi}  &  =-i\hbar
\frac{\mathbf{e}_{\varphi}}{r\sin\theta}\frac{\partial}{\partial\varphi
}=-i\hbar\left(  \frac{\mathbf{e}_{\varphi}}{r\sin\theta}\frac{\partial
}{\partial\varphi}-\mathbf{M}_{\varphi}\right)  .
\end{align}
Now, the mean curvature vectors exhibit themselves respectively in the
brackets, which result from making the derivative\ $-i\hbar\partial_{\xi}$
($\xi=r,\theta,\varphi$) Hermitian operators multiplied, respectively, by the
vector coefficients $\mathbf{e}_{\xi}/H_{\xi}$ with $H_{\xi}$ denoting
Lam\'{e} coefficients for orthogonal curvilinear coordinates, where the
Lam\'{e} coefficients are defined by $d\mathbf{x\equiv}dx\mathbf{e}%
_{x}+dy\mathbf{e}_{y}+dz\mathbf{e}_{z}=\sum H_{\xi}d\xi\mathbf{e}_{\xi}$. For
the spherical polar coordinates ($r,\theta,\varphi$), three Lam\'{e}
coefficients are $H_{r}=1,$ $H_{\theta}=r,$ and $H_{\varphi}=r\sin\theta$. So,
Eq. (\ref{2}) has three ways of decomposition in the following,%
\end{subequations}
\begin{equation}
\mathbf{P=}-i\hbar\nabla_{sp}=\mathbf{e}_{r}P_{r}+\mathbf{\Pi}_{r}%
=\frac{\mathbf{e}_{\theta}}{r}P_{\theta}+\mathbf{\Pi}_{\theta}=\frac
{\mathbf{e}_{\varphi}}{r\sin\theta}P_{\varphi}+\mathbf{\Pi}_{\varphi},
\end{equation}
where, $\mathbf{\Pi}_{r}$, $\mathbf{\Pi}_{\theta}$ and $\mathbf{\Pi}_{\varphi
}$ are so-called the geometric momenta
\cite{GK,homma,ikegami,OK,liu2015,07,liu11,liu13-2,133,liu13-1,135,134} for
the corresponding surfaces, though the last one $\mathbf{M}_{\varphi}=0$ is
trivial,
\begin{subequations}
\begin{align}
\mathbf{\Pi}_{r}  &  \equiv-i\hbar\left(  \mathbf{e}_{\theta}\frac{1}{r}%
\frac{\partial}{\partial\theta}+\mathbf{e}_{\varphi}\frac{1}{r\sin\theta}%
\frac{\partial}{\partial\varphi}+\mathbf{M}_{r}\right)  ,\\
\mathbf{\Pi}_{\theta}  &  \equiv-i\hbar\left(  \mathbf{e}_{r}\frac{\partial
}{\partial r}+\mathbf{e}_{\varphi}\frac{1}{r\sin\theta}\frac{\partial
}{\partial\varphi}+\mathbf{M}_{\theta}\right)  ,\\
\mathbf{\Pi}_{\varphi}  &  \equiv-i\hbar\left(  \mathbf{e}_{r}\frac{\partial
}{\partial r}+\mathbf{e}_{\theta}\frac{1}{r}\frac{\partial}{\partial\theta
}+\mathbf{M}_{\varphi}\right)  .
\end{align}
They are special cases of the general form of the geometric momentum,
\cite{134}
\end{subequations}
\begin{equation}
\mathbf{\Pi}\equiv-i\hbar\left(  \mathbf{r}^{\xi}\partial_{\xi}+\mathbf{M}%
\right)  ,\text{ }or\text{ \ }\mathbf{\Pi}\equiv-i\hbar\left(  \mathbf{r}%
^{\xi}\partial_{\xi}+\frac{\mathbf{M}}{2}\right)  \label{GM}%
\end{equation}
where $\mathbf{M}=M\mathbf{n}$ with $\mathbf{n}$ denoting the unit normal
vector for a surface and $M$ standing for the mean curvature. In the first
equation of (\ref{GM}), the mean curvature $M$ is usually defined by the true
average of the two principal curvatures and ususlly applies for the 2D
surface, whereas the second one uses another convention in which $M$ is
defined as sum of all principal curvatures. In the rest part of this paper, we
will use the latter convention.

\section{A theorem for general cases}

\emph{Theorem:}\ In the $\left(  N+1\right)  D$ Euclidean space $R^{N+1}$, we
can define the usual Cartesian coordinates whose corresponding momentum is
$\mathbf{P=}-i\hbar\nabla_{cart}$ as usual, which can also expressed in terms
of the curvilinear coordinates ($\xi^{0},\xi^{\mu}$), ($\mu=1,2,...N$).
Assuming that the curvilinear coordinates take the form of gaussian normal
coordinates that have a metric that satisfies conditions $g_{00}>0$ does not
depend on\ $\xi^{0}$ and $g_{0\mu}=0$. There is a mean-curvature dependent
decomposition of the momentum in the following,
\begin{equation}
\mathbf{P\equiv}-i\hbar\nabla_{cart}=-i\hbar\left(  \frac{\mathbf{n}}%
{\sqrt{g^{00}}}\frac{\partial}{\partial\xi^{0}}-\frac{\mathbf{M}_{0}}%
{2}\right)  +\mathbf{\Pi}_{0},
\end{equation}
where $-\mathbf{M}_{0}$ is the mean curvature vector, and $\mathbf{\Pi}_{0}$
defines the geometric momentum of the surface $\xi^{0}=const.$ whose unit
normal vector is denoted by $\mathbf{n}$, $\mathbf{\Pi}_{0}\mathbf{=}%
-i\hbar\left(  \mathbf{r}^{\mu}\partial_{\mu}+\frac{\mathbf{M}_{0}}{2}\right)
.$

The proof is straightforward. The coordinate transformation from the cartesian
ones $\mathbf{x\equiv}\left(  x_{1},x_{2},x_{3},...x_{N+1}\right)  $ to the
gaussian normal ones $(\xi^{0},\xi^{\mu})$, ($\mu=1,2,3,...,N$) are,
\begin{equation}
x_{i}=x_{i}(\xi^{0},\xi^{\mu})\text{, and }\xi^{0}=\xi^{0}(\mathbf{x}%
),\xi^{\mu}=\xi^{\mu}(\mathbf{x}).
\end{equation}
The line element\ $d\mathbf{x\cdot}d\mathbf{x}$ is $d\mathbf{x\cdot
}d\mathbf{x}\mathbf{\equiv}dx_{i}dx_{i}=g^{00}\partial_{0}^{2}+g^{\mu\nu
}\partial_{\mu}\partial_{\nu}$, and the the determinant of the metric matrix
$g_{\mu\nu}$ is then $g=\left\vert g_{\mu\nu}\right\vert $. The gradient
operator is in the gaussian normal coordinates,
\begin{equation}
\nabla_{gn}=\frac{\mathbf{n}}{\sqrt{g^{00}}}\partial_{0}+\mathbf{r}^{\mu
}\partial_{\mu}.
\end{equation}
This gradient operator contains no mean curvature. The mean-curvature
dependence becomes evident in quantum momentum in the following.

First, we assume $g^{00}=1$. The canonical momentum operators ($P_{0},P_{\mu}%
$) associated with canonical positions $(\xi^{0},\xi^{\mu})$ are,
respectively, given by,
\begin{equation}
-i\hbar\partial_{0}\rightarrow P_{0}=-i\hbar\frac{1}{g^{1/2}}\partial
_{0}g^{1/2}\text{, }-i\hbar\partial_{\mu}\rightarrow P_{\mu}=-i\hbar\frac
{1}{g^{1/2}}\partial_{\mu}g^{1/2}.
\end{equation}
To note that $\xi^{0}(\mathbf{x})=const.$ forms a surface whose mean curvature
$M_{0}$ is simply, \cite{ikegami}
\begin{equation}
\frac{1}{g^{1/2}}\partial_{0}g^{1/2}=-M_{0}\text{, and }\mathbf{M}_{0}%
=M_{0}\mathbf{n}%
\end{equation}
We have then,
\begin{equation}
\mathbf{P}=-i\hbar\left(  \mathbf{n}\partial_{0}+\mathbf{r}^{\mu}\partial
_{\mu}\right)  =-i\hbar\left(  \mathbf{n}\partial_{0}-\frac{\mathbf{M}_{0}}%
{2}+\mathbf{r}^{\mu}\partial_{\mu}+\frac{\mathbf{M}_{0}}{2}\right)
=-i\hbar\mathbf{n}P_{0}+\mathbf{\Pi}_{0}. \label{fgm}%
\end{equation}
Secondly, for $g^{00}\ $takes any positive values, we can easily prove,%

\begin{equation}
-i\hbar\frac{\partial_{0}}{\sqrt{g^{00}}}\rightarrow P_{0}=-i\hbar\frac
{1}{g^{1/2}}\frac{\partial_{0}}{\sqrt{g^{00}}}g^{1/2}=-i\hbar\left(
\frac{\partial_{0}}{\sqrt{g^{00}}}-\frac{M_{0}}{2}\right)  .
\end{equation}
The decomposition (\ref{fgm}) remains the same.

For clearly see that $\mathbf{\Pi}_{0}$ is really lying on the surface
$\xi^{0}=const.$, we can verify the orthogonal relation $\mathbf{n}%
\cdot\mathbf{\Pi}_{0}+\mathbf{\Pi}_{0}\cdot\mathbf{n}=0$. \cite{134}
\textit{Q.E.D.}

Thus, we understand why there are three mean curvature vectors associated with
$3D$ spherical polar coordinates. This is because any one coordinate is normal
to other two. Moreover, for an $ND$ surface in $R^{N+1}$, any point in the
neighborhood of the surface can be clearly specified by the gaussian normal
coordinates $\mathbf{r(}\xi^{\mu}\mathbf{)}+\xi^{0}\mathbf{n(}\xi^{\mu
}\mathbf{)}$ with $\mathbf{n(}\xi^{\mu}\mathbf{)}$ is the unit normal vector
on point $\xi^{\mu}$ of the surface, and we can define the geometric momentum
on the surface. \cite{liu13-2,134}

\section{Conclusions}

Many come across a fact that the canonical momenta in the orthogonal
curvilinear coordinates are closely related to the mean curvature vectors of
some properly defined curved surfaces, and the mean curvature vectors are
geometric invariant, rather than the Christoffel symbols that give different
values from one set of coordinates to another. We demonstrate that a
decomposition of momentum operator in Gaussian normal coordinates
straightforwardly leads to a natural appearance of the mean curvature vectors.
Once the canonical momentum along the normal becomes hermitian, the remaining
part of the momentum is lying on the surface, which is the geometric momentum
which recently attracts much attention.

\begin{acknowledgments}
This work is financially supported by National Natural Science Foundation of
China under Grant No. 11175063.
\end{acknowledgments}

\end{document}